\documentclass[prd,aps,preprint,nofootinbib,superscriptaddress]{revtex4}
\usepackage{epsfig}
\usepackage{amsmath}
\input{epsf}

\begin{document}

%\preprint{BNL-NT-xxx}\preprint{LBNL-xxxx} \preprint{RBRC-xxx}

\vspace*{2cm}
\title{Drell-Yan Lepton Pair Azimuthal Asymmetry in Hadronic Processes}

\author{Jian Zhou}
%\email{jzhou@lbl.gov}
\affiliation{School of Physics, Shandong University, Jinan, Shandong 250100, China}
\affiliation{ Nuclear Science Division, Lawrence Berkeley National Laboratory, Berkeley, CA 94720}
\author{Feng Yuan}
%\email{fyuan@lbl.gov}
\affiliation{ Nuclear Science Division, Lawrence Berkeley National Laboratory, Berkeley, CA 94720}
\affiliation{RIKEN BNL Research Center, Building 510A, Brookhaven National Laboratory, Upton, NY 11973}
\author{Zuo-Tang Liang}
%\email{liang@sdu.edu.cn}
\affiliation{School of Physics, Shandong University, Jinan, Shandong 250100, China}
%\date{\today}

\begin{abstract}
We study the azimuthal asymmetry ($\cos 2\phi$) in the Drell-Yan lepton pair production in
hadronic scattering processes at moderate transverse momentum region,
taking into account the contributions from the twist-three quark-gluon
correlations from the unpolarized hadrons. The contributions are found to dominate the
asymmetry, and are not power suppressed by $q_\perp/Q$ at small $q_\perp$
where $q_\perp$ and $Q$ are the transverse momentum and invariant
mass of the lepton pair. Accordingly, the Lam-Tung relation will be violated at
this momentum region, and its violation depends on the twist-three functions.
However, at large transverse momentum $q_\perp\sim Q$,
the Lam-Tung relation still holds because all corrections are power suppressed
by $\Lambda^2/q_\perp^2\sim \Lambda^2/Q^2$ where $\Lambda$ is the
typical nonperturbative scale.
\end{abstract}

\maketitle

{\bf 1. Introduction.}
Drell-Yan lepton pair production in hadronic scattering process~\cite{Drell:1970wh}
has been playing a very important role in studying nucleon
structure and QCD dynamics~\cite{Brock:1993sz}, and
is an important complementary to
the deep inelastic scattering studies~\cite{Bloom:1969kc}.
Moreover, the later development on the angular distribution
of lepton pair has laid ground for parton model
and QCD dynamics studies~\cite{Collins:1977iv,Lam:1978pu,{Collins:1978yt}}.
The lepton pair production in hadronic scattering,
\begin{equation}
H_1+H_2\to \gamma^*+X\to \ell^+\ell^-+X \ ,
\end{equation}
comes from the virtual photon decays. At higher energies,
we should also consider the weak boson ($Z^0$) decay
contributions. In this paper, we will limit our discussions
only for the virtual photon decays. An extension to including
$Z^0$ boson decay contributions is straightforward. In the leading
order, virtual photon is produced through quark-antiquark
annihilation process, $q\bar q\to \gamma^*$ in the parton picture~\cite{Drell:1970wh}.
In the rest frame of the lepton pair, we can define two angles~\cite{Collins:1977iv}: one
is the polar angle $\theta$ between one lepton momentum and the hadron;
the azimuthal angle $\phi$ is defined as the angle between the hadronic
plane and the lepton plane. Here and in the following discussions,
we follow the Collins-Soper frame~\cite{Collins:1977iv} to define these angles. Our results
can be translated to other frames too. The general expression for the lepton pair
angular distribution can be written as~\cite{Collins:1977iv},
\begin{eqnarray}
\frac{dN}{d\Omega}=(1+\cos^2\theta)+A_0(\frac{1}{2}-\frac{3}{2}\cos^2\theta)
+A_1\sin2\theta \cos\phi +\frac{A_2}{2}\sin^2\theta\cos2\phi\ .
\end{eqnarray}
It has been argued \cite{Collins:1977iv} that the coefficients $A_0$, $A_1$, and $A_2$ are
all power suppressed at large $Q^2$, where $Q$ is the invariant mass of
the lepton pair: $A_0\sim A_2\sim \langle k_\perp^2\rangle /Q^2$ and $A_1\sim \langle k_\perp\rangle /Q$
where $k_\perp$ is the typical transverse momentum scale in the process~\cite{Collins:1977iv}.
As a result,  the lepton pair angular distribution will be dominated
by $(1+\cos^2\theta)$ in the small transverse momentum region.
These power counting results were generalized to analyze the various
relations between the above coefficients~\cite{Lam:1978pu}. One of the
interesting observations is the so-called Lam-Tung relation~\cite{Lam:1978pu}:
$2\nu-(1-\lambda)=0$, where $\lambda=(2-3A_0)/(2+A_0)$ and $\nu=2A_2/(2+A_0)$.
According to the above power counting results, this relation is
obviously valid because $\lambda=1$ and $\nu=0$ at the leading power.

Finite transverse momentum of the lepton pair ($q_\perp$)
can be generated from gluon radiation from the
leading partonic process, for example, through the quark-antiquark
annihilation channel $q\bar q\to \gamma^* g$. Its contribution
to the lepton pair angular distribution at finite transverse momentum
was found~\cite{Collins:1978yt},
\begin{equation}
\frac{dN}{d\Omega}=\frac{3}{16\pi}\left[\frac{Q^2+\frac{3}{2}q_\perp^2}{Q^2+q_\perp^2}+
\frac{Q^2-\frac{1}{2}q_\perp^2}{Q^2+q_\perp^2}\cos^2\theta+\frac{\frac{1}{2}q_\perp^2}{Q^2+q_\perp^2}
\sin^2\theta\cos 2\phi+\ldots\right]\ ,
\end{equation}
where the $\sin2\theta$ term does not have simple expression
and has been omitted in the above equation. Similar expression
can be derived for the $qg\to \gamma^* q$ channel. Certainly, at finite
transverse momentum, the angular coefficients $A_i$ in Eq.~(2)
are not zero any more. However, they do obey the power
counting rule. For example, the $A_2$ coefficient in Eq.~(2)
from the contribution in Eq.~(3) is power suppressed
by $q_\perp^2/Q^2$ at low transverse momentum.
However, the Lam-Tung relation is still valid for any value of $q_\perp$ from this
contribution. This has been regarded as a simple prediction
from QCD~\cite{Collins:1978yt}. Higher order perturbative
corrections to this relation has been calculated in
the literature~\cite{Mirkes:1994dp}, where it was found that the violation is numerically
very small. These studies have motivated many experimental investigations
of the lepton pair angular distributions in hadronic scattering~\cite{Conway:1989fs}.
In particular, the $\pi$ induced fixed target experiment found large $\cos 2\phi$
azimuthal asymmetry, which is difficult to understand~\cite{Conway:1989fs,{Chiappetta:1986yg}}.
The intrinsic transverse momentum effects have been proposed to
explain these effects~\cite{Boer:1999mm}, which however is limited to small transverse momentum
whereas the experimental data are in both small and moderate transverse momentum
region.

Meanwhile, at low transverse momentum, there exist large logarithms
in terms of $\alpha_s^{n}\ln^{2n} (Q^2/q_\perp^2)$ from the fixed order perturbative
calculations~\cite{Dokshitzer:1978dr,Parisi:1979se,Collins:1981uw,Collins:1984kg,{Boer:2006eq}}.
Resummation of these larger logs has been formulated
for the leading contribution from the partonic contribution, especially
for the Drell-Yan lepton pair and vector boson production in hadronic
scattering~\cite{Collins:1984kg}. This corresponds to the leading term in the lepton pair
angular distribution $(1+\cos^2\theta)$ in Eq.~(2). Because the rest terms
($A_i$) are power suppressed in the limit $q_\perp/Q\ll1$, it has been
difficult to follow the Collins-Soper-Sterman
resummation method~\cite{Chiappetta:1986yg}. Recently, there have been much efforts to study
the soft gluon resummation for these higher order terms~\cite{{Berger:2007si}}, and
hopefully these developments will lead to a final solution to this
issue, especially following the original Collins-Soper-Sterman
formalism.

In this paper, we study the lepton pair azimuthal asymmetry, in particular for
the $A_2$ coefficient in Eq.~(2) from different perspective.
We are interested in its behavior at the moderate
transverse momentum region $\Lambda_{QCD}\ll q_\perp\ll Q$.
At this region, there are two large momentum
scales $q_\perp$ and $Q$. The contribution to $A_2$ from Eq.~(3) is power suppressed
by $q_\perp^2/Q^2$ as we mentioned above.
However, from the following calculations we find that
there exit contributions from the twist-three quark-gluon correlation functions from
both incident hadrons,
which are not power suppressed by $q_\perp^2/Q^2$, instead by $\Lambda^2/q_\perp^2$
where $\Lambda$ is the typical nonperturbative scale. These contribution
will dominate $A_2$ coefficient at the moderate transverse momentum region,
depending on the relative strength of $q_\perp^2/Q^2$ and $\Lambda^2/q_\perp^2$~\footnote{From
power counting point of view, at this particular order, the new contribution will dominate in the
region of $\Lambda^2\ll q_\perp^2\ll \Lambda Q$ whereas the contribution from
Eq.~(3) will dominate in the region of $\Lambda Q\ll q_\perp^2\ll Q^2$.}.
More importantly, because they are not power suppressed by $q_\perp^2/Q^2$, the soft
gluon resummation can be performed following the classical Collins-Soper-Sterman
approach, and the resummation effect will be similar to the leading term
of $(1+\cos^2\theta)$ (see for example, the similar study in~\cite{Idilbi:2004vb}).
We notice that another higher-twist effect from only one side of the incoming
hadrons has also been studied in the literature~\cite{{qiu-fac},Braun:2000kw},
which are different from our calculations below.

{\bf 2. Twist-three times twist-three contributions to
the lepton pair azimuthal asymmetry.}
From the general analysis of twist-three functions of the unpolarized
hadrons~\cite{Ellis:1982cd,{Jaffe:1991ra}},
we find the only twist-three function is $T_F^{(\sigma)}(x,y)$
which is equivalent to $E(x,y)$ studied in the literature~\cite{Jaffe:1991ra,{Zhou:2008mz}}.
It is defined as
\begin{eqnarray}
T_F^{(\sigma)}(x_1,x_2)=\int \frac{dy_1^- dy_2^-}{4 \pi}
e^{iy_2^-(x_2-x_1)P^++ iy_1^-x_1P^+} \langle P| \bar{\psi}(0^-)
\sigma ^{+\mu} g {F^+}_{\mu}(y_2^-) \psi(y_1^-) |P \rangle \ ,
\end{eqnarray}
where $\mu$ is a transverse index, the sums over color and spin indices are implicit, $|P \rangle$
denotes the unpolarized hadron state with momentum
$P=(P^+,0^-,0_\perp)$ and $P^\pm=(P^0\pm P^z)/\sqrt{2}$,
$\psi $ is the quark field,
and $ F_{+\mu}$ the gluon field tensor, and the gauge link has been
suppressed. This correlation is a chiral-odd function, and
can generate transverse polarized Hyperon production in
unpolarized hadronic collisions~\cite{Zhou:2008fb}. Because
of this chirality property, we have to introduce two correlation
functions from both incoming hadrons.

To calculate its contribution, we follow
the procedure outlined in~\cite{Ellis:1982cd,qiu}, and recent developments
for the similar calculations~\cite{new,Eguchi:2006qz,jqvy}.
In the collinear factorization framework, a general
factorization formula for the contributions from the above correlation
functions can be written as,
\begin{equation}
\frac{d\sigma}{d^4qd\Omega}=\sum_q \int\frac{dxdx'}{x}\frac{dzdz'}{z}
T_{F,q}^{(\sigma)}(x,x')T_{F,\bar q}^{(\sigma)}(z,z') {\cal H}(x,x';z,z';Q^2,q_\perp) \ ,
\end{equation}
where $T_{F,q}^{(\sigma)}(x,x')$ is the correlation function associated with
the quark from hadron $H_1$ and $T_{F,\bar q}^{(\sigma)}(z,z')$ for
the antiquark from hadron $H_2$, ${\cal H}$ is the hard part and can
be calculated from perturbative partonic process. This factorization
formula follows (as a conjecture) earlier general arguments for the higher-twist contributions
to the hadronic cross sections~\cite{Ellis:1982cd,{qiu-fac}}.
It will be very important to have a rigorous proof for this particular contribution
as written in Eq.~(5): the higher-twist effects coming from both sides of incoming hadrons,
which is beyond the situations considered in~\cite{qiu-fac}.
Because of the higher-twist nature, it is always much more involved
to calculate their contributions than those for the leading-twist contributions
like Eq.~(3). However, recent developments~\cite{new,Eguchi:2006qz,jqvy}
have laid solid ground and useful technique to carry out those calculations.
In particular, the two variables in $T_F^{(\sigma)}$ will be fixed by taking pole contributions,
or equivalently by calculating the imaginary part of the interference of the scattering
amplitudes. For example, in the above equation $x'$ will be equal to
$x$ depending on a soft or hard pole contribution~\cite{new,Eguchi:2006qz,jqvy}.
In this paper, we will follow the
procedure developed in~\cite{new,Eguchi:2006qz,jqvy}
to calculate the hard part in Eq.~(5).

First, we notice that  the hard partonic part is separately gauge invariant summing up
all possible diagrams. Therefore,  we can carry out the calculations of these contributions
with either $A^+$ or $A_\perp$ field connecting the hard and soft
parts. A particular example has been given in~\cite{{Eguchi:2006qz}} for similar calculations.
In our calculations, we find that it is more convenient to work the $A_\perp$ part, and construct
the field tensor accordingly. However, it has been known that the individual diagrams
associated with the $A_\perp$ field depend on the boundary condition,
although the final results of all diagrams contributions do not~\cite{Ji:2002aa,{Brodsky:2002cx}}.
Further calculations show that the retarded boundary condition $A_\perp(y^-=-\infty)=0$
will greatly simplify the derivations of Eq.~(5). In the following, we will choose this
boundary condition for the gauge field $A_\perp$ from both hadrons.
Under this boundary condition, on the other hand, we have to take into account the contributions
from the operator with partial derivative on the
quark field $\left(\bar\psi\partial_\perp \psi\right)$. This is because, this
matrix element can be related to the quark-gluon correlation function defined
in Eq.~(5),
\begin{eqnarray}
\int \frac{dy^-}{4\pi} e^{ixP^+ y^-} \langle P |
\bar{\psi}(0) \sigma^{+\mu}
i\partial_{\perp\mu} \psi_\alpha(y^-) | P \rangle=
T_F^{(\sigma)}(x)\equiv T_F^{(\sigma)}(x,x) \ ,
\end{eqnarray}
with retarded boundary condition for the gauge field~\cite{future}. Therefore, for a complete
calculations, we have to calculate the diagrams associated with the operators
$\left(\bar\psi\partial_\perp \psi\right)$, together with that of
$\left(\bar\psi A_\perp\psi\right)$~\cite{Ellis:1982cd}.
For the $A_\perp$ contribution, we can further deduce its contribution to
that of $T_F^{(\sigma)}(x_1.x_2)$. For example, under the same
boundary condition for the gauge field $A_\perp$, we can write
\begin{eqnarray}
&& \int \frac{dy^-dy_1^-}{4 \pi }P^+ e^{ix_1P^+ y^-}
e^{i(x-x_1)P^+y_1^-}\langle PS | \bar{\psi}(0^-)
\sigma^{+\mu} gA_{\perp\mu}(y_1^-)
\psi(y^-) | PS \rangle\nonumber\\
&=&\frac{i}{x-x_1+i\epsilon}T_F^{(\sigma)}(x,x_1) \ ,
\end{eqnarray}
where the pole structure in the second line comes from the
partial integral and the $i\epsilon$ prescription
depends on the retarded boundary condition we are using.
If we choose different boundary condition, this
prescription shall change accordingly~\cite{future}. We have also checked
that the above procedure can reproduce all previous results
~\cite{new,Eguchi:2006qz,jqvy}.

\begin{figure}[t]
\begin{center}
\includegraphics[width=10cm]{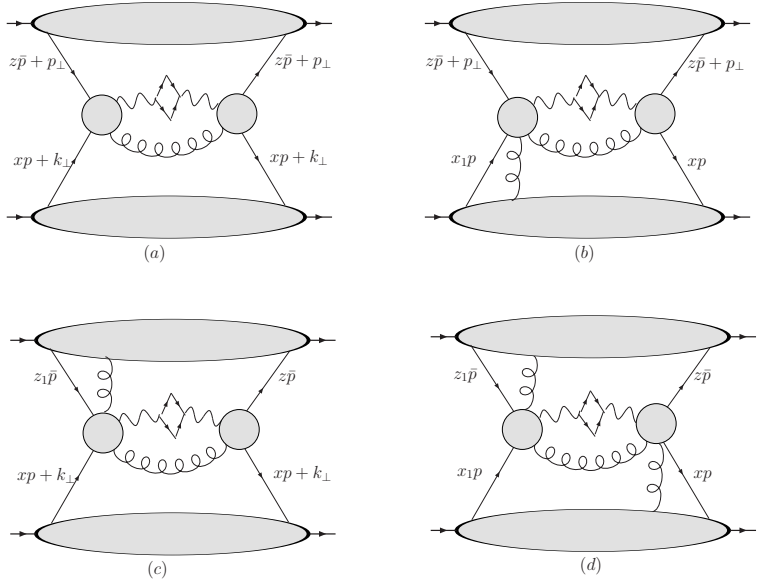}
\end{center}
\vskip -0.4cm \caption{Generic Feynman diagrams for the
twist-three times twist-three contributions to the
lepton pair azimuthal asymmetry in hadronic scattering process:
(a) $\partial_\perp$ contribution from both sides of hadrons:
(b-c) $\partial_\perp$ and $A_\perp$ from each side;
(d) $A_\perp$ from both sides.} \label{fig1}
\end{figure}

Following the above arguments, we plot the generic Feynman diagrams
contributions in Fig.~1, where (a) is the contribution
from $\partial_\perp$ operators from both sides of hadrons;
(b-c) are those diagrams with $\partial_\perp$ and $A_\perp$ on
either side; (d) for $A_\perp$ from both sides. There are four diagrams
for Fig.~1(a), 12 diagrams for Fig.~1(b) and (c) respectively, and 208 diagrams
for Fig.~1(d).
As we mentioned above, in this paper we are interested in the cross section contributions in the
moderate transverse momentum region $q_\perp\ll Q$.
In carrying out these calculations, we will utilize the power counting
method. We will only keep the leading power contributions and neglect all
higher power corrections in terms of $q_\perp/Q$. The full expressions
of our results will be presented in a separate publication.
The advantage to use the retarded boundary condition
is that we find that at the leading power, the contributions from the diagrams
of Fig.~1(d) are either power suppressed or canceled out between soft
and hard poles~\cite{future}.
We are left with the contributions from Fig.~1(a-c) only, which are relatively
easier to work out. In the limit of $q_\perp\ll Q$, the final result is
\begin{eqnarray}
\frac{d \sigma}{d^4q d\Omega}&=&\sigma_0\sin^2\theta \cos(2\phi)\frac{2}{q_\perp^4} \frac{\alpha_s}{2\pi^2}
\sum_q e_q^2
\nonumber \\&&
\int \frac{dx}{x} \frac{dz}{z}  \left \{ A {T}_{F,\bar{q}}^{(\sigma)}(z,z) \delta(\hat{\xi}-1)
+ \bar{A} T_{F,q}^{(\sigma)}(x,x) \delta(\xi-1) \right .\
\nonumber \\&&  \ \ \ \  \ \ \ \ \ \ \ \ \  \left .\
+ 2 C_F \delta(\xi-1) \delta(\hat{\xi}-1)T_{F,q}^{(\sigma)}(x,x) {T}_{F,\bar{q}}^{(\sigma)}(z,z)
{\rm ln} \frac{Q^2}{q_\perp^2} \right \} \ ,
\end{eqnarray}
where $\sigma_0=\alpha_{em}^2/6SQ^2$, $\xi=x_0/x$, $\hat \xi=z_0/z$ with $x_0=\frac{Q}{\sqrt{S}}e^y$ and $z_0=\frac{Q}{\sqrt{S}} e^{-y}$,
and $y$ is the rapidity of the lepton pair in the center of mass frame of incoming two hadrons and
$S$ is the hadronic center of mass energy square. The coefficient $A$ is defined as
\begin{eqnarray}
{A}=\frac{1}{2 N_c} \left \{ \left [ x \frac{\partial}{\partial x} {T}_{F,q}^{(\sigma)}(x,x) \right ]
2 {\xi}+ {T}_{F,q}^{(\sigma)}(x,x)\frac{2 {\xi}({\xi}-2)}{(1-{\xi})_+} \right \}
+\frac{C_A}{2} {T}_{F,q}^{(\sigma)}(x,x_0)\frac{2}{(1-{\xi})_+} \  ,
\end{eqnarray}
and similar expression holds for $\bar A$ in the above equation.
These diagrams (Fig.~1) also contribute to other terms including $A_0$ and $A_1$
in Eq.~(2), but they are all power suppressed by $q_\perp^2/Q^2$.
From the above equations, we can see that this contribution to $A_2$ coefficient
is not power suppressed by $q_\perp^2/Q^2$ at moderate transverse
momentum region. It is the same order as the $(1+\cos^2\theta)$ term
in this power counting. Of course, it is suppressed by $\Lambda^2/q_\perp^2$
because of the higher-twist nature. This can also be seen from the above expression.

More importantly, the above result can also be reproduced
by a transverse momentum dependent (TMD) factorization formalism~\cite{Brodsky:2002cx,Collins:2002kn,Ji:2002aa,Boer:2003cm,JiMaYu04,ColMet04,Mulders:1995dh}
with the so-called Boer-Mulders function from both hadrons~\cite{Boer:1997nt,{Metz:2008ib}} at large
transverse momentum which has been calculated in~\cite{Zhou:2008fb}.
This demonstrates that in the intermediate transverse
momentum region, for this part contribution,
the twist-three times twist-three collinear factorization approach and
the TMD factorization approach are consistent for the $\cos(2\phi)$
azimuthal asymmetry in the unpolarized Drell-Yan processes. This is a nontrivial
demonstration, because it goes beyond previous examples studied in the literature~\cite{jqvy}
where the twist-three effect from only one side of the incoming hadrons was considered.
Because of this consistency,
the energy evolution equation~\cite{Collins:1981uw} (the Collins-Soper evolution equation)
can be derived for this contribution, and the soft gluon resummation
can be accordingly performed. This will significantly change the relative sizes
of this contribution and the contribution from Eq.~(3) to the $\cos 2\phi$ asymmetry.
We will leave a detailed study in a separate publication~\cite{future}.

{\bf 3. Conclusion.}
We have the following results for the angular distribution
of the Drell-Yan lepton pair production in hadronic reactions,
\begin{itemize}
\item At moderate transverse momentum,
$A_2$ is in order of 1, $A_0$ is power suppressed by $q_\perp^2/Q^2$. As
a result, the Lam-Tung relation will be violated because $\lambda$ is 1 whereas
$\nu$ is order of 1. Of course this violation will depend on the sizes of the
twist-three correlation function $T_F^{(\sigma)}$ from both incoming hadrons.
Furthermore, soft gluon resummation
will not change the power counting result for $A_2$, because the
leading order TMD factorization leads to the same resummation pattern similar to
that discussed in~\cite{Collins:1984kg}, which is important to
understand the angular distributions of the lepton pair at this momentum
region~\cite{Chiappetta:1986yg}.

\item At large transverse momentum $q_\perp\sim Q$, however,
both $A_0$ and $A_2$ are in order of 1, and they are dominated by
the contribution from the unpolarized quark and antiquark annihilation contribution
Eq.~(3). The contributions we calculated this paper are suppressed
by $\Lambda^2/q_\perp^2\sim \Lambda^2/Q^2$.
Because of this, the  Lam-Tung relation will be valid again.
\end{itemize}

In summary, we have investigated the higher-twist effects
to the Drell-Yan lepton pair angular distributions in hadron-hadron
scattering processes. We found that the twist-three times twist-three
contributions to the $\cos2\phi$ azimuthal
asymmetry contribution are not power suppressed by $q_\perp^2/Q^2$,
rather by $\Lambda^2/q_\perp^2$ at the moderate transverse momentum.
We further argue that this part of contribution
will not be affected by the soft gluon resummation effects, and the Lam-Tung
relation will be modified at small and moderate transverse momentum.
It will be interested to compare our predictions with the
experimental data~\cite{Conway:1989fs} and check the phenomenological importance
of our results. It will also be interested to extend to other processes
like the $\cos 2\phi$ asymmetry in semi-inclusive hadron
production in deep inelastic scattering and back-to-back two hadron production
in $e^+e^-$ annihilation processes where the similar effects shall play
very important roles.

We thank Jianwei Qiu and Werner Vogelsang for
interesting discussions. This work was supported in part by the U.S. Department of Energy
under contract DE-AC02-05CH11231 and the National Natural Science
Foundation of China under the approval No. 10525523. We are grateful
to RIKEN, Brookhaven National Laboratory and the U.S. Department of
Energy (contract number DE-AC02-98CH10886) for providing the
facilities essential for the completion of this work. J.Z. is
partially supported by China Scholarship Council.

\end {document}